\documentclass[runningheads]{llncs}
\usepackage{graphicx}
\usepackage[]{threeparttable}
\usepackage{multirow}
\usepackage{makecell}
\begin{document}
\title{Understanding the Hoarding Behaviors during the COVID-19 Pandemic using Large Scale Social Media Data}
%
%
\author{Xupin Zhang\and
Hanjia Lyu\and
Jiebo Luo }
\authorrunning{X. Zhang et al.}
%
\institute{University of Rochester, Rochester NY 14627, USA \\
\email{xzhang72@u.rochester.edu}\\
\email{hlyu5@ur.rochester.edu}\\
\email{jluo@cs.rochester.edu}}
\maketitle              
\begin{abstract}
The COVID-19 pandemic has affected people's lives around the world on an unprecedented scale. We intend to investigate hoarding behaviors in response to the pandemic using large-scale social media data. First, we collect hoarding-related tweets shortly after the outbreak of the coronavirus. Next, we analyze the hoarding and anti-hoarding patterns of over 42,000 unique Twitter users in the United States from March 1 to April 30, 2020, and dissect the hoarding-related tweets by age, gender, and geographic location. We find the percentage of females in both hoarding and anti-hoarding groups is higher than that of the general Twitter user population. Furthermore, using topic modeling, we investigate the opinions expressed towards the hoarding behavior by categorizing these topics according to demographic and geographic groups. We also calculate the anxiety scores for the hoarding and anti-hoarding related tweets using a lexical approach. By comparing their anxiety scores with the baseline Twitter anxiety score, we reveal further insights. The LIWC anxiety mean for the hoarding-related tweets is significantly higher than the baseline Twitter anxiety mean. Interestingly, beer has the highest calculated anxiety score compared to other hoarded items mentioned in the tweets.

\end{abstract}
\section{Introduction}
The study of consumer behaviors examines the processes that individuals or groups experience when they purchase or use products in order to meet their needs~\cite{arndt2004urge}. Shopping patterns, as a part of consumer behaviors, refer to the typical way in which people buy goods or services. Historically, marketing research has relied on questionnaires and focus groups to measure or analyze shopping patterns~\cite{sim2002singapore}. The COVID-19 pandemic provides a golden opportunity to study consumer behaviors during a massive crisis and discover insights that would otherwise be difficult to attain during normal times.


During the COVID-19 pandemic, social media has acted as a double-edged sword: while it is a rich source for obtaining useful information concerning the pandemic, it also shapes the fears. For instance, when posts of panic-buying (e.g., toilet paper, hand sanitizer) proliferate on social media platforms, people might make panic purchases after seeing such posts. It is unclear how hoarding behaviors expressed on social media differ from hoarding behaviors documented before. In addition, it is not known whether tweeted shopping behaviors adequately reflect the shift in actual panic buying behaviors since the COVID-19 outbreak. We build our hypotheses on the notion that panic buying behavior differs across gender, age and family status during unstable times~\cite{hori2014run}. In addition, we hypothesize that geographic location impacts what items to hoard (e.g., urban vs. rural; coastal vs. inland states). By analyzing the emotions, attitudes, and opinions of adults within a three-month period, we hope to have a more comprehensive understanding of why and what consumers hoarded.

This study uses Twitter data during the COVID-19 pandemic to investigate hoarding-related tweets. To our knowledge, this is the first study to examine hoarding behaviors using social media. In particular, this study makes two main contributions: 

\begin{itemize}
    \item We demonstrate that the hoarding behaviors during the COVID-19 pandemic differ across age, gender, and geographic location.
    \item We further conduct a content analysis using topic modeling~\cite{blei2003latent} and analyze the anxiety scores of the tweets using a lexical approach~\cite{pennebaker2015development}. 
\end{itemize}

\section{Related Work}
Hoarding was originally termed “compulsive hoarding behavior”. Hoarding behavior is the acquisition of and failure to discard a large number of possessions that seem to be useless or of limited value~\cite{frost1996cognitive}. Researchers have applied this definition to studying hoarding behavior over the years~\cite{tolin2008economic}. However, most of the studies documented clinical hoarding behaviors. The hoarding behavior we discuss in this paper is more related to the behavior known as panic buying, rather than hoarding disorder.

Online behaviors, such as the ones detected on Twitter are representative of real human behaviors. Bekafigo and McBride~\cite{bekafigo2013tweets} found that online political activists are identical to offline activists, which means that offline political activists are frequently the ones who have high levels of political participation on Twitter. Another study~\cite{proserpio2016psychology} examined the relationship between psychological well-being and unemployment using 1.2 billion Twitter posts concerning job loss or job gain from 2010 to 2015. The study showed that social media posts are better at explaining the causes and consequences of unemployment than traditional economic models. In addition, researchers have been studying psychological language to analyze social economic issues using Twitter. For instance, Eichstaedt et al.~\cite{eichstaedt2015psychological} examined the relationship between psychological language on Twitter and heart disease mortality. The authors regressed heart disease mortality on the dictionary and topic language variables, holding income and education constant, and found the correlation between psychological language and heart disease mortality are significant.

\section{Methods}
\subsection{Data Collection and Preprocessing}
Twitter data is collected from Crimson Hexagon, which is a social media analytic platform. We use hashtags (\#panicbuying, \#panicpurchasing, \#hoarding) and keywords (“panic buying”, “panic purchasing”, “panic hoarding”, “hoarding”, “panic-buying", “panic-hoarding", “panic shopping”, “panic buy”, “panic purchase”) to extract tweets. A total of 1,421,954 tweets related to hoarding behaviors from March 1 to April 30, 2020 are downloaded. Features include the tweet ID, the screen name of the author, tweet content, the date the tweet was posted, and the location of the author. There are 980,185 unique Twitter users. Since our study focuses on the consumers living in the U.S., we drop the data of the Twitter users who are living outside the U.S. 

To characterize the Twitter users, we use the Twitter API to obtain information from the Twitter users’ personal profiles. The Twitter user’s name, user ID, location information, profile photo, and personal description are downloaded. As suggested by An and Weber~\cite{an2016greysanatomy} and Lyu et al.~\cite{lyu2020sense}, we apply Face++\footnote{https://www.faceplusplus.com/}, an image recognition platform, to infer attributes including the number of faces, gender, and age. People may use photos of themselves or group photos. We keep the profile images where there is only one intelligible face. We also eliminate the tweets not written in English. 

To obtain more insights into the geo-locations of the tweet authors, we classify them according to the population density and whether they are located in coastal or inland states. For the population density, we use a Python package\footnote{https://pypi.org/project/uszipcode/} called \textit{uszipcode} to classify the tweet author into three groups – urban (over 3,000 people per square mile) , suburban (1,000 – 3,000 people per square mile), and rural (less than 1,000 people per square area).

In the end, our dataset includes 42,839 unique US Twitter users with four inferred features including age, gender, population density, and whether the consumer lives in coastal or inland states.

\subsection{Categorization of Groups: Mentions Hoarding versus Anti-Hoarding}

It is possible that the tweets that mentioned “hoarding” or “panic buying” are the ones that actually ask people to stop hoarding or panic buying. To find these tweets and their authors, we lemmatize the tweet contents and apply a simple rule-based method. If the word “stop”, “don’t panic”, “no need” or “no panic” is in the tweets, then this tweet and the author are labeled as “anti-hoarding”. 10.8\% of consumers are labeled as “anti-hoarding”. To check the validity of this method, we sample 100 tweets from the set that we label as “anti-hoarding” and 100 tweets from the set which we label as “hoarding”, and manually label them. Of 100 tweets that are labeled as “anti-hoarding”, 98\% express the idea of anti-hoarding, and 2\% do not. Of 100 tweets that are labeled as “hoarding”, 84\% do not express the idea of anti-hoarding while 16\% do. This suggests that the method is  sufficient to find the consumers who are against hoarding. Therefore, we separate the dataset into two groups: tweets mentioning hoarding or panic buying (Hoarding Group, HG) and Anti-Hoarding Group (AHG). There are 38,207 (89.2\%) users in HG and 4,632 (10.8\%) users in AHG. Descriptive statistics for HG and AHG are reported in Table~\ref{table1}. 

\begin{table*}[htbp]
\scriptsize
\centering
\caption{Characteristics of HG/AHG}
\label{table1}
\begin{tabular}{lllll}
\hline
                     & \multicolumn{2}{l}{Hoarding} & \multicolumn{2}{l}{Anti-Hoarding} \\
Variable             & Frequency    & Percentage    & Frequency       & Percentage      \\
\hline
Age                  &              &               &                 &                 \\
18-35                & 21,946       & 57.4          & 2,548           & 55.0            \\
36-54                & 11,380       & 29.8          & 1,458           & 31.5            \\
\textgreater{}55     & 4,881        & 12.8          & 626             & 13.5            \\
\hline
Gender               &              &               &                 &                 \\
Male                 & 17,432       & 45.6          & 2,098           & 45.3            \\
Female               & 20,775       & 54.4          & 2,534           & 54.7            \\
\hline
Population   Density &              &               &                 &                 \\
Urban                & 30,777       & 80.6          & 3,706           & 80.0            \\
Suburban             & 3,776        & 9.9           & 509             & 11.0            \\
Rural                & 3,654        & 9.6           & 417             & 9.0             \\
\hline
Coastal   and Inland &              &               &                 &                 \\
Coastal              & 32,111       & 84.0          & 3,924           & 84.7            \\
Inland               & 6,096        & 16.0          & 708             & 15.3            \\ \hline
\end{tabular}
\end{table*}

\section{Results}

\subsection{Demographic Differences}
We perform a goodness-of-fit test and find sufficient evidence to conclude that the age distributions of HG and AHG are statistically different ($p<.001$). The authors of AHG are older than the authors of HG. For HG, 
there are more younger hoarders (57.4\%) who are between 18 and 35 years old. As the college students fall into this group and Twitter is heavily used by college students~\cite{junco2007connecting}, this may suggest that college students actively expressed their hoarding behaviors when they stay at home after their colleges abruptly shut down since mid-March. Similarly, middle-aged adults who are between 36 and 54 years old actively post hoarding related tweets more than usual while they work from home. 

Comparing with the gender distribution of U.S. Twitter users as of January 2020\footnote{https://www.statista.com/statistics/678794/united-states-twitter-gender-distribution/}, we find that the gender distributions of HG/AHG demonstrate a significant female bias: The percentage of the female hoarders (54.4\%) is higher than the female authors in the general Twitter population (43.8\%). Similarly, the percentage of female anti-hoarders (54.7\%) is higher than the female authors in general Twitter users (43.8\%). 

After performing the goodness-of-fit test, we find that the population distributions are statistically different between AHG and HG ($p<.001$). The result is in line with the study of Hori and Iwamoto~\cite{hori2014run}, which also shows that a majority of hoarders are from urban areas.

In addition, we separate the consumers into coastal and inland groups based on their self-reported locations. After performing the proportion z test, we find that the proportion of the Twitter users living in the coastal states of AHG is statistically higher than that of the Twitter users living in the inland states ($p<.001$). 
The consumers living in coastal states tend to post tweets that tell people to stop hoarding or panic buying. One possible reason is that residents of the coastal states are relatively more affluent or educated than those of inland states.\footnote{https://www.cbsnews.com/news/american-migration-rich-move-to-coasts-poor-to-the-heartland/} There are also more online shoppers than mall-shoppers in the coastal states.\footnote{https://www.washingtonpost.com/business/economy/no-the-rest-of-america-is-not-online-shopping-like-you-are/2017/03/13/2812a9c8-05ab-11e7-b1e9-a05d3c21f7cf\_story.html} Therefore, consumers that live in the coastal states may buy products online instead of going to the stores to hoard.

\subsection{Tweet Content Analysis using Topic Modelling}

To find out what the consumers focus on, we use Latent Dirichlet Allocation (LDA) to identify probable topics of the tweets of HG. Table~\ref{tab:topics} shows the 3 topics generated by LDA and the top 10 keywords of each topic. We manually assign a topic name for each topic. As we can see from the table, the consumers of HG are mainly concerned with toilet paper, food and shortage.

\begin{table*}[!ht]
\scriptsize
\centering

\begin{threeparttable}
\caption{Topics generated by the LDA model using the HG tweets. }
\begin{tabular}{l | l | l}
\hline
\textbf{Group} & \textbf{Topics} & \textbf{Top 10 Topic Words} \\
 \hline
	
\multirow{3}{*}{\textbf{HG}} & Toilet paper & paper toilet panic buy people else know everyone take come\\
& Food & panic people hoard need food go get buy grocery store\\
& Shortage & panic go home buy people COVID start need last shortage\\
\hline
\end{tabular}
\label{tab:topics}

\end{threeparttable}

\end{table*}


	




\begin{figure}[htbp]
    \centering
    \includegraphics[width = 0.6\linewidth]{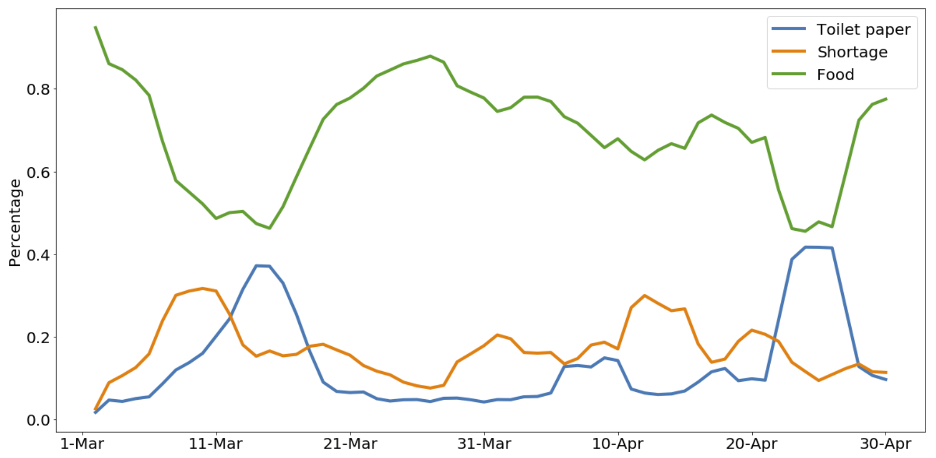}
    \caption{Temporal change in the proportion of the dominant topics of HG from March 1 to April 30, 2020.}
    \label{fig:topic_trend}
\end{figure}

The LDA model assigns weights to the three topics of each individual tweet. The topic with the highest weight is the dominant topic of this tweet. To better understand the characteristics of the consumers in HG and their opinions, we analyze the proportions of dominant topics across demographics and the geo-locations. Fig.~\ref{fig:topic_trend} shows the smoothed temporal changes of the proportions of dominant topics of HG. Food was always the primary topic within HG during this time period. The proportions of the other two topics decreased as the pandemic developed. Discussions about toilet paper were at peak around the mid of March. This trend is consistent with the actual purchasing behaviors documented by the news, which states that the panic behavior moved from toilet paper to meat starting from the mid of March\footnote{https://www.post-gazette.com/opinion/editorials/2020/05/09/No-need-to-hoard-Panic-buying-moves-from-toilet-paper-to-meat/stories/202005090012}. Another interesting finding is that the discussions about toilet paper increased a little after April 20 and decreased in around 10 day. This is probably due to the restock of the toilet paper. According to the data Reuters reported from NCSolutions, about 73 percent of US stores were out of toilet paper on April 12; the number went down to 48 percent by April 19\footnote{https://www.vox.com/the-goods/2020/5/13/21256144/toilet-paper-companies-workers-coronavirus-pandemic}.

\subsubsection{Age}
In all three age groups, the major topic was food, suggesting that food was the priority concern of the consumers of all ages. By performing the goodness-of-fit test, we find that the distributions of the topics of the three age groups are statistically different ($p<.001$). The “18-35” age group tends to have a more diverse distribution of topics: 56.0\% talk about food, 25.2\% talk about toilet paper and 18.7\% talk about shortage. However, the topics are more consistent in the other two age groups. 70.2\% and 73.9\% consumers of the “36-55” and the “over 56” age groups, respectively, talk about food.

\subsubsection{Gender}
Food was the dominant topic across both genders. According to the goodness-of-fit test, the distributions of topics across the gender groups are statistically different ($p<.001$). The distribution of topics among females is more diverse. 58.1\% females talk about food, 23.7\% talk about toilet paper, and the final 18.2\% talk about shortage. Males also pay more attention to food. 67.8\% of males talk about food, 14.4\% are concerned with toilet paper. The proportion of the males that talk about shortage is only 17.8\%, which is relatively lower than that of females.

\subsubsection{Population Density}
With respect to the population density, there are statistical differences ($p<.001$) between the distributions of the topics of the consumers living in the urban, suburban and rural areas. However, the differences are small. Across all three groups, the major topic is food. By performing the goodness-of-fit test, we find there is sufficient evidence ($p<.001$) to conclude that the distributions of topics of the consumers living in the coastal states and inland states are different. However, similar to the analysis of population density, the differences are not big. Within both groups, the major topic is food.

\subsection{Sentiment Analysis using a Lexical Approach}
LIWC is a lexical approach to sentiment analysis using its internal lexicons~\cite{pennebaker2015development}. LIWC has also been applied in previous studies to conduct sentiment analysis using social media data~\cite{chen2020eyes}. Although LIWC has several sentiment categories, we have selected four major sentiment categories (positive emotion, negative emotion, anger, and anxiety) for our analysis. The LIWC2015 document guide presents the baseline anxiety mean scores for six genres of texts~\cite{pennebaker2015development}\footnote{Grand means refer to the unweighted means of the six genres of texts (Blogs, Expressive writing, Novels, Natural Speech, NY times, Twitter).}. We use these as a baseline to compare the linguistic and sentiment scores for each category. We refer to the score of Twitter text as “LIWC Twitter mean” and the average score of six genres as “LIWC Grand mean” in the results, respectively.  In addition, we have also calculated the LIWC mean for the hoarding related tweets. We referred to this mean as “LIWC Calculated Anxiety Hoarding Mean” and “LIWC Calculated Anxiety Anti-Hoarding Mean” respectively\footnote{Hoarding and Anti-Hoarding calculated means are the LIWC results using our collected hoarding related tweets.}.

Table~\ref{tab:liwc} shows the summary profiles for hoarding and anti-hoarding related tweets. T test shows that hoarding tweets have a significantly higher future-focused and past focused score ($p<.05$) than anti-hoarding tweets ($p<.05$), while anti-hoarding tweets have a significantly higher present focus than hoarding tweets ($p<.001$). To better understand the difference, we follow Gunsch et al.~\cite{gunsch2000differential} to analyze four pronoun scores and time orientation scores together. The anti-hoarding tweets have more other references (“she/he”, “they”) than hoarding tweets, while they have fewer self-references such as “we”, we can infer that the tweets of anti-hoarding tweets focus on past and present actions of others more than the baseline tweets do.

\begin{table*}[h]
\centering
\scriptsize
\begin{threeparttable}
        \caption{LIWC results.}
    \begin{tabular}{c|c|c|c|c|c}
    \hline
        \makecell{Category} & \makecell{Example} & \makecell{LIWC \\Twitter Mean} & \makecell{LIWC \\Grand Mean \\(SD)} & \makecell{LIWC \\Calculated\\ Hoarding \\Mean} & \makecell{LIWC \\Calculated \\Anti-Hoarding\\ Mean}\\
     \hline
        \textbf{Linguistic Processes} & & & & &\\
        \hline
        Personal pronouns & I, them, her & 9.02 & 9.95 (3.02) & 5.94 & 6.21\\
        \hline
        $1^{st}$ person singular & I & 4.75 & 4.99 (2.46) & \textbf{1.72} & 1.53\\
        \hline
        $1^{st}$ persons plural & we & 0.74 & 0.72 (0.83) & 0.75 & \textbf{0.99}\\
        \hline
        $2^{nd}$ person & you & 2.41 & 1.70 (1.35) & 1.98 & \textbf{2.33} \\
        \hline
        $3^{rd}$ person singular & she, him & 0.64 & 1.88 (1.53) & 0.28 & \textbf{0.30}\\
        \hline
        $3^{rd}$ persons plural & they, their & 0.47 & 0.66 (0.60) & \textbf{1.28} & 0.81\\
        \hline
        \textbf{Psychological Processes} & & & & &\\
        \hline
        
        Affect Processes & happy, cried & 7.67&	5.57 (1.99)&5.90&\textbf{6.72}\\
        \hline
        
          Positive emotion&love, nice&5.48&3.67 (1.63)&1.77&\textbf{2.41}\\
        \hline
        
         Negative emotion&hurt, ugly&2.14&1.84 (1.09)&4.09&\textbf{4.28}\\
        \hline
        
         Anxiety&worried, fearful&0.24&0.31 (0.32)&\textbf{2.67}&2.57\\
        \hline
        
            Anger&hate, kill&0.75&0.54 (0.59)&0.67&\textbf{0.63}\\
        \hline

            Sadness&crying, sad&0.43&0.41 (0.40)&0.28&\textbf{0.55}\\
        \hline
         \textbf{Time Orientations} & & & & &\\
        \hline
          Past focus&ago, did&2.81&4.64(2.06)&2.11&\textbf{1.45}\\
          \hline
  Present focus&today, is&11.74&9.96(2.80)&10.84&\textbf{12.02}\\
  \hline
  Future focus&may, will&1.60&1.42(0.90)&\textbf{1.37}&1.01\\
  \hline

    \end{tabular}

    \label{tab:liwc}

    \begin{tablenotes}[flushleft]
\small
\item{{Note: Grand means refers to the unweighted means of the six genres of texts (Blogs, Expressive writing, Novels, Natural Speech, NY times, Twitter). Twitter means refer to the LIWC output using 35, 269 Twitter author’s Twitter posts~\cite{pennebaker2015development}. Hoarding and Anti-Hoarding calculated means are the LIWC results using our collected tweets.}}
\end{tablenotes}
\end{threeparttable}

\end{table*}

As Table~\ref{tab:liwc} shows,  the calculated means for the negative emotions of hoarding and anti-hoarding tweets are both much higher than the LIWC Grand mean and LIWC Twitter mean. This suggests that hoarding-related tweets show a higher negative emotion level compared to the LIWC averages and LIWC general tweets averages. As for anxiety, the sentiment scores for both hoarding and anti-hoarding tweets are much higher than the LIWC means. This indicates that both hoarding and anti-hoarding tweets express a high level of anxiety, which is in line with previous research showing that anxiety is one of the core reasons to hoard~\cite{darrat2016impulse}.

\begin{figure}[htbp!]
    \centering
    \includegraphics[width = 0.71\linewidth]{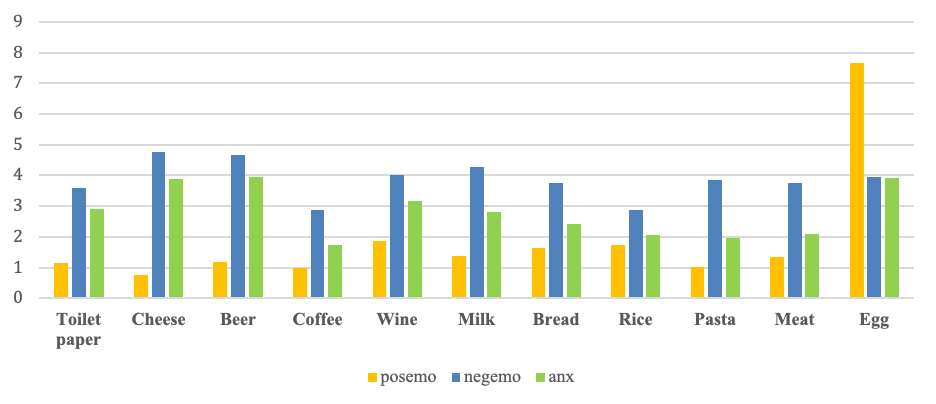}
    \caption{Food-related hoarding tweets by emotion.}
    \label{fig*:food}
\end{figure}

To cover as many food-related keywords as possible, the Food Pyramid~\cite{marcus2013culinary} and the What We Eat in America (WWEIA) list~\cite{agricultural2010we} are used as reference lists to extract the items from the food hoarding related tweets. We analyze the food items that were mentioned in at least 100 hoarding related tweets. As can be seen from Fig.~\ref{fig*:food}, we find that beer 
has a significantly higher calculated anxiety score and negative emotion score than the other food items ($p<.05$) using the t test. This suggests that individuals hoarded for beer to reduce anxiety during this period of time. The results are consistent with the study of Thomas et al.~\cite{thomas2003drinking} that shows that individuals drink alcohol to relieve anxiety. In addition, we find that egg has a significantly higher positive emotion score than the other food items ($p<.05$) using the t test. 

\begin{figure}[htbp]
    \centering
    \includegraphics[width=.71\linewidth]{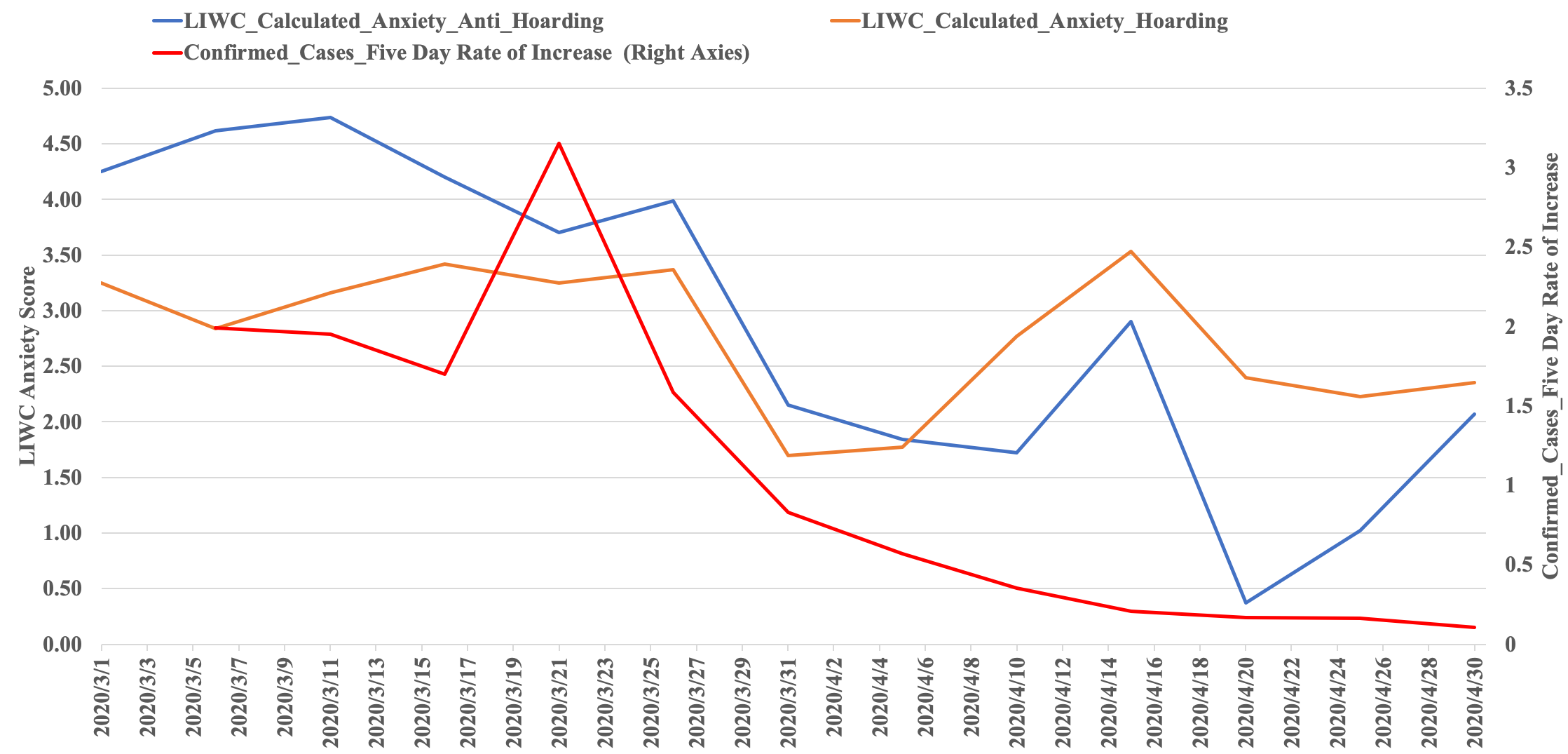}
    \caption{LIWC anxiety levels of the hoarding and anti-hoarding tweets from March 1 to April 30, 2020.}
    \label{fig*:liwc_pic}
    \vspace{-0.55cm}
\end{figure}

Fig.~\ref{fig*:liwc_pic} shows the trend of the LIWC calculated anxiety scores and the growth rate of COVID-19 cases in the US from March 1 to April 30, 2020. The calculated anxiety means for hoarding and anti-hoarding tweets are presented. It shows that both calculated means are above the LIWC Twitter mean and LIWC Grand mean for anxiety shown in Table 3 (0.24 and 0.31, respectively). This suggests that anxiety levels are higher than the expected LIWC mean. Both LIWC-calculated hoarding anxiety score and anti-hoarding anxiety score experienced a sharp increase from mid- to late March. Similarly, the COVID-19 5-day infection rate followed a sharp increase and decrease in the same period. We find that the anxiety level of the hoarding tweets peaked before the peak of new COVID-19 cases.

\section{Conclusion and Future Work}
In this paper, we investigate the hoarding or panic buying behavior during the outbreak of the COVID-19 pandemic by analyzing the hoarding related tweets. We first apply a rule-based method to separate the tweets and their authors into two groups on the basis of whether the tweets indicate hoarding behaviors or express the opinion against hoarding. We find there are significant differences between the consumers of these two groups across age, gender, the population density of their locations, and whether they live in the coastal states. We then apply an LDA model to investigate the topics of the hoarding consumers. We find that the tweets of the hoarding group focus on food, toilet papers and shortage to various degrees depending on the demographics. Furthermore, we analyze the sentiment of the tweets using LIWC2015. The hoarding-related tweets show more anxiety. Interestingly, beer has the highest calculated anxiety score compared to other hoarded items mentioned in tweets. In the future, we plan to explore additional consumer behavior theories using such data to better compare the consumer behaviors online and offline.




%
%
%
\bibliographystyle{splncs04}
\bibliography{mybibliography}

\end{document}